\documentclass[preprint2]{aastex}

\usepackage{graphicx}
\DeclareGraphicsExtensions{.ps}  
\shorttitle{\indent \def Magnetic reconnection and chromospheric evaporation} \shortauthors{Tian et al.}

\begin{document}

\title{Imaging and spectroscopic observations of magnetic reconnection and chromospheric evaporation in a solar flare}

\author{Hui Tian\altaffilmark{1}, Gang Li\altaffilmark{2}, Katharine K. Reeves\altaffilmark{1}, John C. Raymond\altaffilmark{1}, Fan Guo\altaffilmark{3}, Wei Liu\altaffilmark{4}, Bin Chen\altaffilmark{1}, Nicholas A. Murphy\altaffilmark{1}}
\altaffiltext{1}{Harvard-Smithsonian Center for Astrophysics, 60 Garden Street, Cambridge, MA 02138; hui.tian@cfa.harvard.edu}
\altaffiltext{2}{University of Alabama in Huntsville, Huntsville, AL} \altaffiltext{3}{Los Alamos National Lab, Los Alamos, NM}
\altaffiltext{4}{Stanford-Lockheed Institute for Space Research, CA}

\begin{abstract}
Magnetic reconnection is believed to be the dominant energy release mechanism in solar flares. The standard flare model predicts both downward
and upward outflow plasmas with speeds close to the coronal Alfv\'{e}n speed. Yet, spectroscopic observations of such outflows, especially the
downflows, are extremely rare. With observations of the newly launched Interface Region Imaging Spectrograph (IRIS), we report the detection of
greatly redshifted ($\sim$125~km~s$^{-1}$ along line of sight) Fe~{\sc{xxi}}~1354.08\AA{} emission line with a $\sim$100~km~s$^{-1}$ nonthermal
width at the reconnection site of a flare. The redshifted Fe~{\sc{xxi}} feature coincides spatially with the loop-top X-Ray source observed by
the Reuven Ramaty High Energy Solar Spectroscopic Imager (RHESSI). We interpret this large redshift as the signature of downward-moving
reconnection outflow/hot retracting loops. Imaging observations from both IRIS and the Atmospheric Imaging Assembly (AIA) onboard the Solar
Dynamics Observatory (SDO) also reveal the eruption and reconnection processes. Fast downward-propagating blobs along these loops are
also found from cool emission lines (e.g., Si~{\sc{iv}}, O~{\sc{iv}}, C~{\sc{ii}},
Mg~{\sc{ii}}) and images of AIA and IRIS. Furthermore, the entire Fe~{\sc{xxi}} line is blueshifted by $\sim$260~km~s$^{-1}$ at the loop
footpoints, where the cool lines mentioned above all exhibit obvious redshift, a result that is consistent with the scenario of chromospheric
evaporation induced by downward-propagating nonthermal electrons from the reconnection site.
\end{abstract}


\section{Introduction}
Magnetic reconnection is believed to be the dominant energy release mechanism during solar flares. Early observations of flares have led to the
development of the standard flare model, in which the frequently observed H$\alpha$~ribbons, post-flare loops, hard X-ray sources at the loop
footpoints and the loop top can be explained as consequences of magnetic reconnection \citep[e.g.,][]{Magara1996}. After the mid-1990s,
high-resolution observations at multiple wavelengths have greatly enhanced our understanding of the reconnecting current sheets
\citep[e.g.,][]{Liu2010} and chromospheric evaporation \citep[e.g.,][]{Milligan2009}, which are also part of the standard flare model. For
recent reviews of theories and observations of magnetic reconnection in solar flares, we refer to \cite{Benz2008} and \cite{Fletcher2011}.

There have been observations of plasma inflows in flare reconnections \cite[e.g.,][]{Yokoyama2001}. Upward outflows are often identified as
impulsive ejection of plasmoids. High-speed downflows immediately above the post-eruption flare arcades (supra arcade downflows) were first
presented by \cite{McKenzie1999}, but their nature is subject to debate \citep{Savage2012a,Cassak2013,LiuR2013,Guo2014}. Shrinking loops have
been seen in several imaging observations \citep{Reeves2008,Savage2010,Liu2013}. Spectroscopic observations of fast reconnection outflows
\citep{Wang2007,Hara2011}, especially the downflows, are extremely rare.

We report the first detection of a large redshift in Fe~{\sc{xxi}}~1354.08\AA{} (formation temperature $\sim$10 MK) with the Interface Region
Imaging Spectrograph \citep[IRIS,][]{DePontieu2014} in a C1.6 flare peaked at 17:19 UT on 2014 April 19. Based on observations of IRIS, the
Atmospheric Imaging Assembly \citep[AIA,][]{Lemen2012} onboard the Solar Dynamics Observatory (SDO), and the Reuven Ramaty High Energy Solar
Spectroscopic Imager \citep[RHESSI,][]{Lin2002}, we interpret this large redshift as the reconnection downflow. Greatly blueshifted
Fe~{\sc{xxi}}~features have also been detected in the evaporation flow. Furthermore, inflows, upward-moving outflows, and fast
downward-moving blobs have all been detected. Thus, rather than fragmented pieces of the evidences of reconnection identified in many previous
flare observations, our observations provide a more complete picture of flare reconnection.

\section{Observations}

\begin{figure*}
\centering {\includegraphics[width=\textwidth]{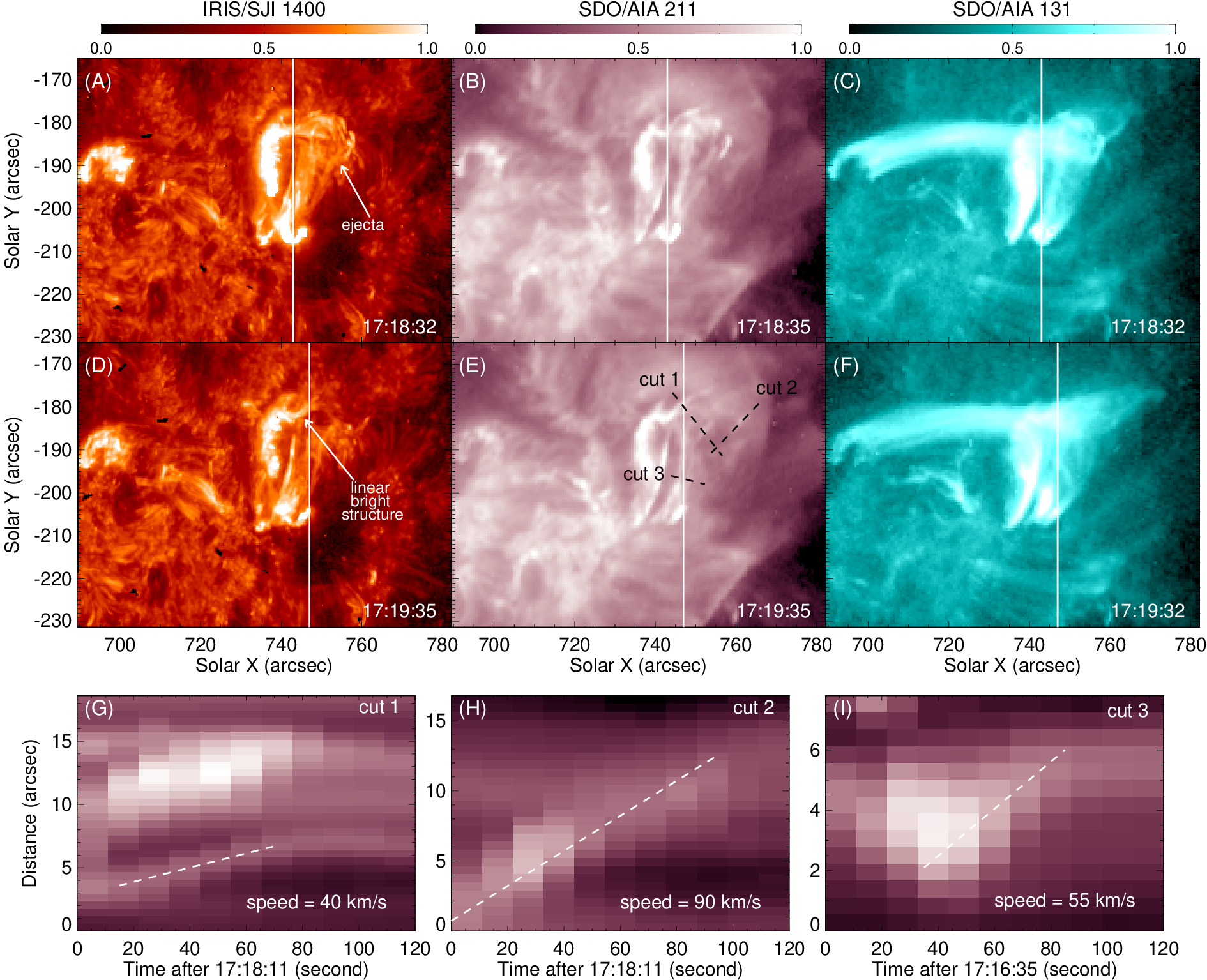}} \caption{ (A)-(C): IRIS/SJI~1400\AA{}, SDO/AIA~211\AA{}~and 131\AA{}~images taken around
17:18:32. The white line in each panel indicates the slit location at the corresponding time. The white arrow in (A) points to the structure ejected later.
(D)-(F): Images taken around 17:19:35. A linear bright structure is indicated by the white arrow in (D). Three cuts shown as dashed lines in (E) are
used to construct the space-time maps in (G)--(I). For each cut, the maximum distance represents the end with the labeled cut number in (E). Two movies
showing the IRIS and AIA observations are available online. } \label{fig.1}
\end{figure*}

The IRIS observation was taken from 15:09 UT to 22:20 UT on 2014 April 19. There were several large coarse 64-step rasters in this observation
and the flare of interest (SOL2014-04-19T17:19 UT) was observed in the fourth raster. The pointing coordinate was (695$^{\prime\prime}$,
-221$^{\prime\prime}$). The data was summed by 2 onboard both spectrally and spatially, leading to a spatial pixel size of 0.33$^{\prime\prime}$
and a spectral dispersion of $\sim$0.026 \AA{} per pixel in the far ultraviolet wavelength bands. The cadence of the spectral observation was
31 seconds, with an exposure time of 30 seconds. Slit-jaw images (SJI) in the 1400\AA{} filter were taken at a cadence of 31 seconds. Dark
current subtraction, flat field and geometrical corrections have been applied in the level 2 data used here. The fiducial lines were used to
achieve a coalignment between the SJI and different spectral windows.

\begin{figure*}
\centering {\includegraphics[width=\textwidth]{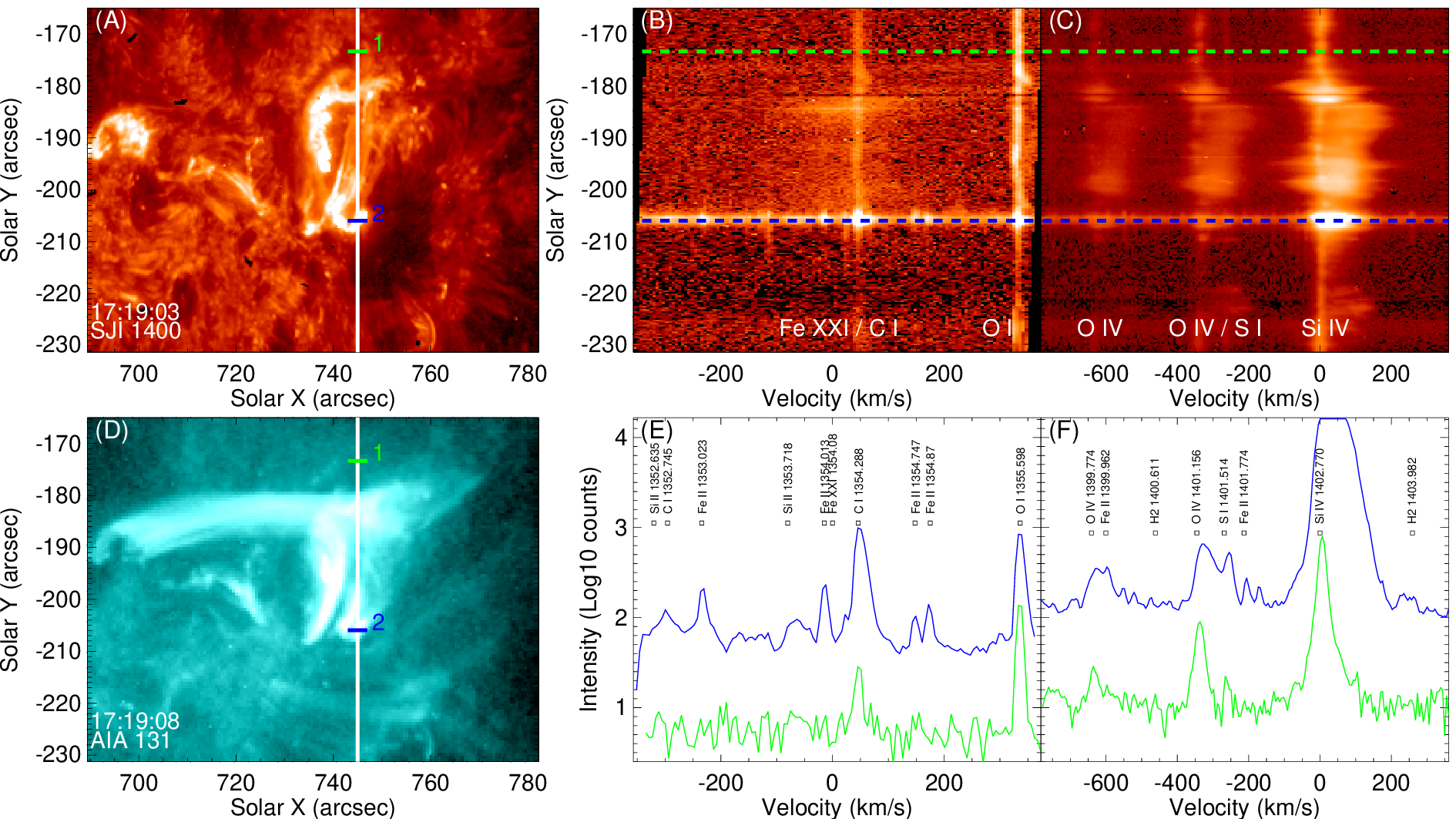}} \caption{ (A)-(C): SJI~1400\AA{} image and detector images of the
Fe~{\sc{xxi}}~1354.08\AA{}~and Si~{\sc{iv}}~1402.77\AA{}~spectral windows at 17:19:03 UT. The zero velocities are set to the rest wavelengths of
these two lines in the two spectral windows. Two locations are labelled as "1" and "2" and indicated by the two horizontal lines with
different colors. (D): SDO/AIA~131\AA{}~image at 17:19:08 UT. (E)-(F): Spectral line profiles at the two locations. } \label{fig.2}
\end{figure*}

\begin{figure*}
\centering {\includegraphics[width=\textwidth]{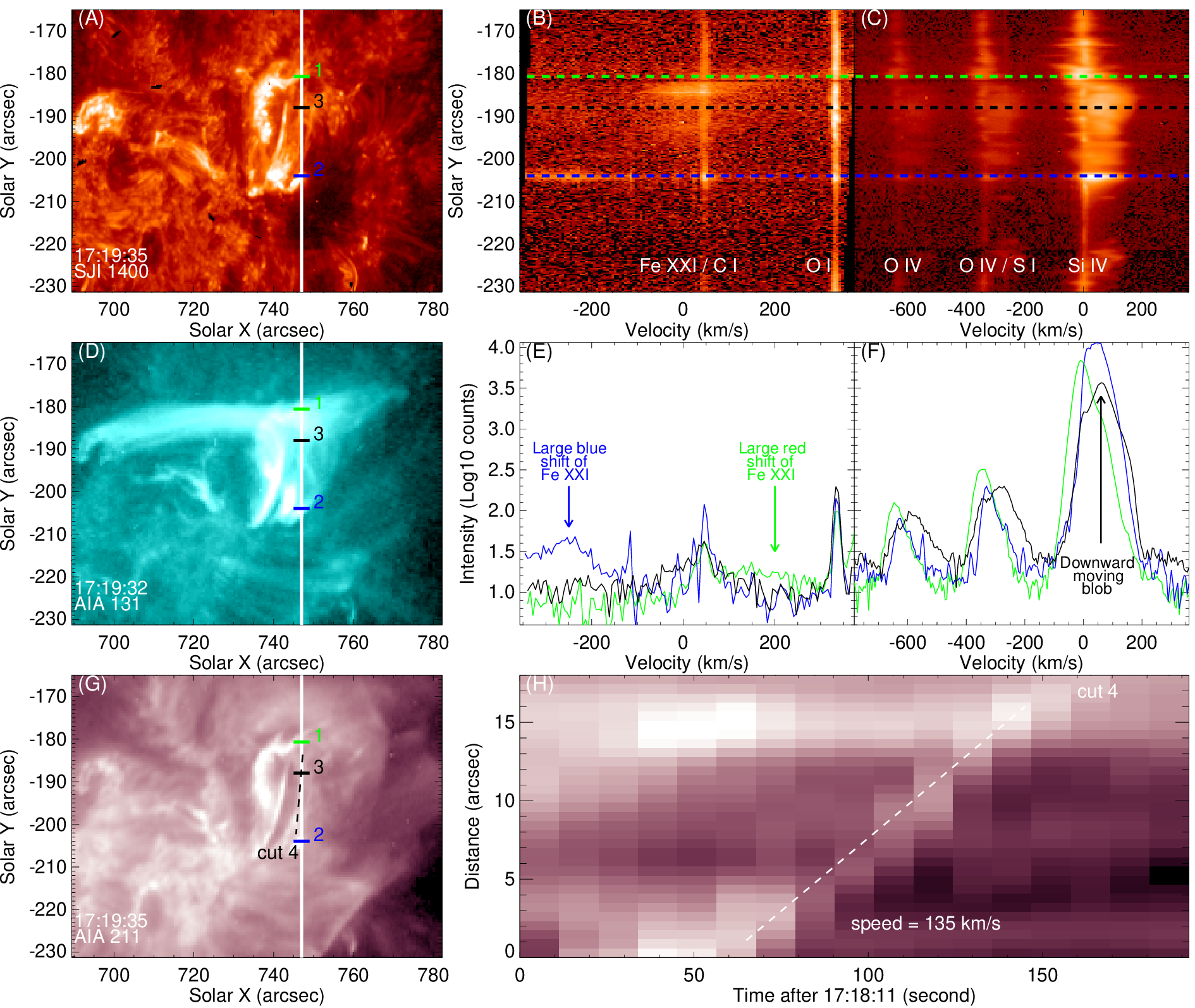}} \caption{ (A)-(C): SJI~1400\AA{} image and detector images of the
Fe~{\sc{xxi}}~1354.08\AA{}~and Si~{\sc{iv}}~1402.77\AA{}~spectral windows at 17:19:35 UT. (D): SDO/AIA~131\AA{}~image at 17:19:32 UT. (E)-(F):
Spectral line profiles at the three locations marked by the horizontal lines with different colors. (G): SDO/AIA~211\AA{}~image at 17:19:35 UT. A cut shown as the dashed line is
used to construct the space-time map in (H).}
\label{fig.3}
\end{figure*}

\begin{figure*}
\centering {\includegraphics[width=\textwidth]{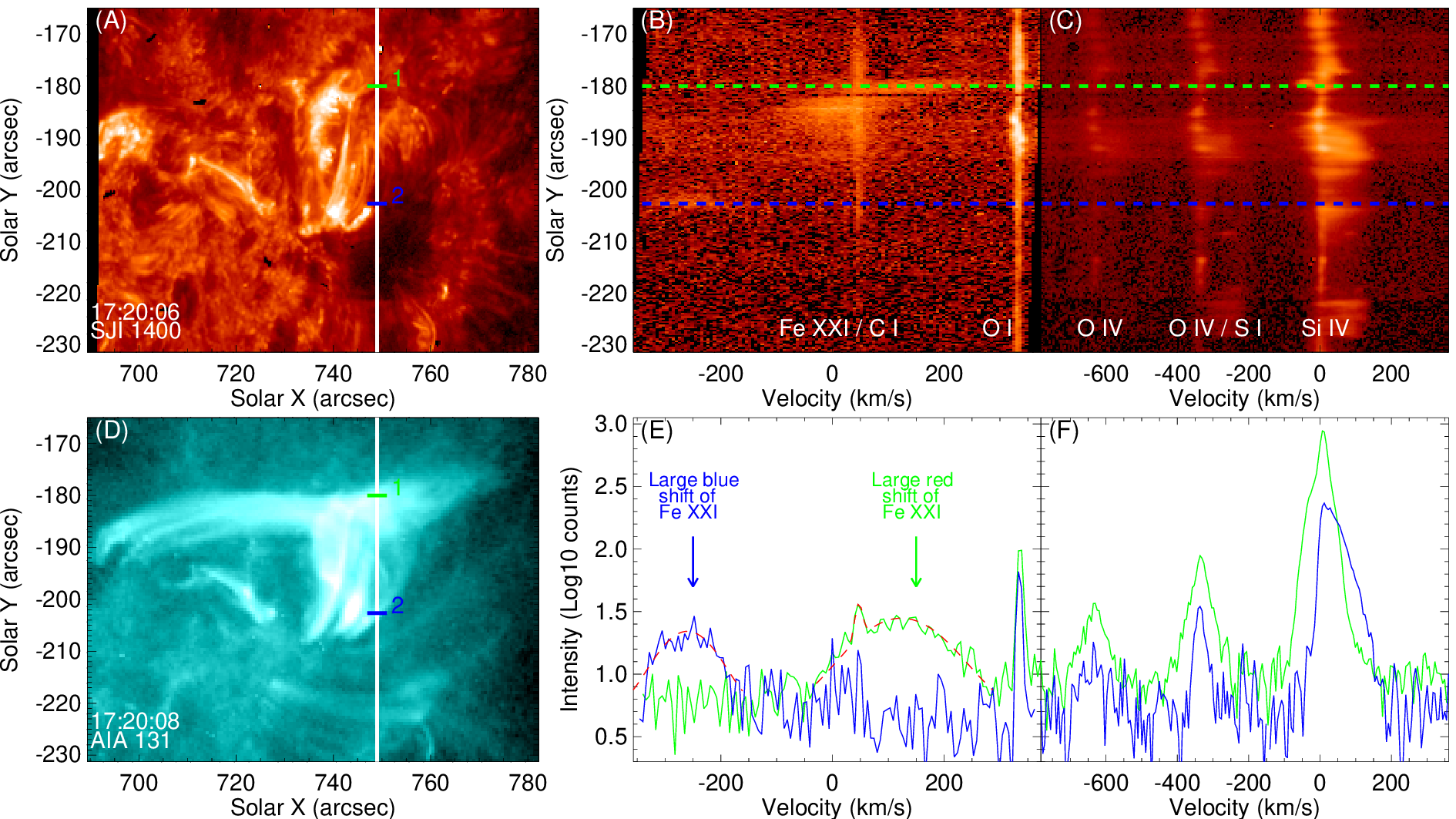}} \caption{ (A)-(C): SJI~1400\AA{} image and detector images of the
Fe~{\sc{xxi}}~1354.08\AA{}~and Si~{\sc{iv}}~1402.77\AA{}~spectral windows at 17:20:06 UT. (D): SDO/AIA~131\AA{}~image at 17:20:08 UT. (E)-(F):
Spectral line profiles at the two locations marked by the short horizontal lines in (A) and (D) and the dashed lines in (B) and (C). The red
dashed lines represent the Gaussian fitting to the redshifted and blueshifted Fe~{\sc{xxi}}~1354.08\AA{}~features. } \label{fig.4}
\end{figure*}

Figure~\ref{fig.1} shows some context images taken by IRIS and AIA. Emission in the IRIS/SJI 1400\AA{} filter is dominated by the strong
Si~{\sc{iv}}~1393.76\AA{}/1402.77\AA{}~lines (formed around 10$^{4.9}$K) and UV continuum \citep[formed slightly above the temperature
minimum,][]{Vernazza1981}. The AIA 1600\AA{}~images and SJI 1400\AA{}~images are used for the coalignment of the images taken by the two
instruments. The emission pattern and plasma motions are similar in the AIA 171\AA{}, 193\AA{}, 211\AA{}, 335\AA{}, 304\AA{} and 1600\AA{}
passbands. We thus only present the 211\AA{}~images here. This passband is dominated by emission from ions formed at both coronal
(Fe~{\sc{xiv}}) and transition region (e.g., O~{\sc{iv}}, O~{\sc{v}}) temperatures \citep{DelZanna2011}. The two hot AIA passbands, 131\AA{} and
94\AA{}, sample emission mainly from the 10 MK and 6 MK plasma respectively. Figures~\ref{fig.2}--\ref{fig.4} present the IRIS/SJI 1400\AA{}
images, corresponding AIA 131\AA{} images, and the IRIS spectra in the Fe~{\sc{xxi}}~1354.08\AA{}~and Si~{\sc{iv}}~1402.77\AA{}~windows at three
different times. We present the time sequence of IRIS observation (including SJI 1400\AA{} images and several spectral windows) and the
corresponding AIA images in an online movie. The full-cadence (12 seconds) movie of AIA 211\AA{} is also available online.

From the online movies we can see that the flare loops brightened at $\sim$17:13:20, $\sim$17:15:55 and $\sim$17:17:30 before the flare peak
($\sim$17:19:00), suggesting a multiple-episode energy release behavior of this flare. In this paper we mainly focus on various types of flows
observed in the third episode starting from $\sim$17:17:30, which were clearly observed from both imaging and spectroscopic observations.

\section{Flows around the reconnection site}

\subsection{Imaging observations}

Observations of IRIS/SJI 1400\AA{}~and AIA 211\AA{} (Figure~\ref{fig.1}) reveal a process consistent with the flare reconnection scenario. Sides
of loops moved towards each other from 17:18:11 to 17:19:35. A bright structure (the ejecta marked in Figure~\ref{fig.1}(A), presumably a flux
rope or plasmoid) can be identified near the coordinates (755$^{\prime\prime}$, -185$^{\prime\prime}$) around 17:18:32. Its disappearance at
17:19:03 is likely related to magnetic reconnection, which permits the field reconfiguration that allows the structure to be expelled. At
17:19:35 a linear bright structure (Figure~\ref{fig.1}(D)) above the post-reconnection loops in the 1400\AA{}~and 211\AA{}~images can be seen.
This resembles the tip of newly formed loops (cusp). It could also be the lower end of the current sheet.

By plotting the space-time map of AIA~211\AA{} (Figure~\ref{fig.1}(G)) along cut 1 (Figure~\ref{fig.1}(E)) we have inferred the speed of
reconnection inflow from the movement of the southern part of the reconnecting loops and found a value of $\sim$40~km~s$^{-1}$. Inflow at the
northern part is not obvious, probably due to projection effect or asymmetric reconnection inflows. Propagation of the ejecta is barely visible
after the eruption, but we do find a bright outward moving feature from AIA~211\AA{} (Figure~\ref{fig.1}(E), cut 2). This feature seems to be
located at the southern boundary of the ejected structure and its speed is estimated to be $\sim$90~km~s$^{-1}$ (Figure~\ref{fig.1}(H)), likely
the lower end of the speed of the upward-moving reconnection outflow.

\subsection{Red shift of Fe~{\sc{xxi}}~1354.08\AA{}}

There are several neutral and singly ionized lines in the spectral windows shown in Figures~\ref{fig.2}--\ref{fig.4} \citep[e.g.,][]{Mason1986}.
However, most of these lines are only visible at the loop footpoints (e.g., location 2 in Figure~\ref{fig.2}). The major blend of the
Fe~{\sc{xxi}}~1354.08\AA{} line is C~{\sc{i}}~1354.288\AA{}, which is usually strong even outside the flare regions (e.g., location 1 in
Figure~\ref{fig.2}).

Significant enhancement at the red side of the Fe~{\sc{xxi}}~1354.08\AA{}~rest position can be identified at the bright linear structure around
17:19:35 (Figure~\ref{fig.3}, location 1). Clearly, this emission feature cannot be explained by the two narrow Fe~{\sc{ii}}~lines marked in
Figure~\ref{fig.2}(E). This enhancement is due to the greatly redshifted Fe~{\sc{xxi}}~1354.08\AA{}~emission. The AIA 131\AA{}~passband also
mainly samples emission from the Fe~{\sc{xxi}}~ion and we clearly see a very bright emission feature at location 1, supporting our argument that
this enhancement is caused by the Fe~{\sc{xxi}}~emission. The enhanced emission extends to the nearby O~{\sc{i}}~line, suggesting redshift of
Fe~{\sc{xxi}}~by up to $\sim$300~km~s$^{-1}$. Images of SJI 1400\AA{}~and AIA 131\AA{}~suggest that location 1 is probably at the lower end of a
current sheet or close to the cusp of the newly reconnected hot loops. So the large redshift of Fe~{\sc{xxi}} is most likely a signature of the
downward-moving reconnection outflow or hot retracting loops. The main blend of the O~{\sc{iv}}~1399.774\AA{}~line is
Fe~{\sc{ii}}~1399.962\AA{}. The main blend of the O~{\sc{iv}}~1401.156\AA{}~line is S~{\sc{i}}~1401.514\AA{}. We take the ratio of the intensities of the two
O~{\sc{iv}}~lines derived from multiple Gaussian fit and obtain a density around log ({\it N}$_{e}$/cm$^{-3}$)=11.0 at location 1 based on CHIANTI v7.1 \citep{Landi2013}.

At 17:20:06 the slit moved to the west and similar spectral features persisted (Figure~\ref{fig.4}). The greatly redshifted Fe~{\sc{xxi}}~emission at
location 1 became much stronger at this time. Applying a double Gaussian fit to this
broad Fe~{\sc{xxi}}~feature and the embedded C~{\sc{i}}~emission feature yields a centroid redshift of $\sim$125~km~s$^{-1}$ and a nonthermal
width (1/e width) of $\sim$96~km~s$^{-1}$. The nonthermal width is calculated based on an assumed kinetic temperature of 10 MK
($\sim$54~km~s$^{-1}$ thermal broadening).

After 17:20:06, the Fe~{\sc{xxi}}~downflow became weaker and the velocity decreased. Presumably, the reconnection site moved to a higher height
and the reconnection also weakened over time. At about 17:21:39 the centroid of the Fe~{\sc{xxi}}~line profile still had a redshift of
$\sim$50~km~s$^{-1}$ and the location of this hot downflow seems to coincide with the loop-top X-ray source observed with RHESSI
(Figure~\ref{fig.5}(A)-(B)). Unfortunately RHESSI was in orbit night before 17:21:30.

The large redshift of Fe~{\sc{xxi}}~is probably the first unambiguous detection of hotter-than-10 MK reconnection downflow/hot retracting loops
in flares using combined imaging and spectroscopic observations. \cite{Wang2007} detected significant enhancement in both wings of an
Fe~{\sc{xix}}~line and interpreted them as reconnection outflows in a flare. The Fe~{\sc{xix}}~line is mainly sensitive to $\sim$8 MK plasma.
\cite{Hara2011} observed a loop-top hot source in the Fe~{\sc{xxiii}} and Fe~{\sc{xxiv}} lines formed at $\sim$12 MK. But these lines exhibited
a redshift of only $\sim$30~km~s$^{-1}$, making it difficult to properly interpret. \cite{Innes2003} detected enhanced blue wing emission of
Fe~{\sc{xxi}}~indicative of plasma flows up to $\sim$1000~km~s$^{-1}$ in a limb flare, and provided the sunward reconnection outflow as one of
the three possible explanations.

The standard flare model also includes a termination shock (TS) resulting from the interaction between the reconnection downflow and the flare
loops. The TS may play a crucial role in electron acceleration and the hardening of the flare spectra above $\sim$300 keV
\citep{Li2013,Kong2013}. \cite{Warmuth2009} and \cite{Gao2014} interpreted type-II radio bursts without frequency drift as a signature of the
TS. Direct evidences of the TS should include a velocity discontinuity, which could potentially be observed by IRIS. In our event, no clues
about the TS can be obtained since we have only observed one redshift. Note that the Fe~{\sc{xxi}}~spectral window used in our observation extends
only to $\sim$360~km~s$^{-1}$ on the red side of Fe~{\sc{xxi}}, so downflows with speeds higher than $\sim$360~km~s$^{-1}$ could not be
detected. Future IRIS observations with a more appropriate choice of the spectral window will be helpful to detect TS.

The co-spatial redshifted Fe~{\sc{xxi}}~feature and the RHESSI soft X-ray source suggest electron heating in the reconnection outflow/hot
retracting loops \cite[e.g.,][]{Liu2013} or at the TS \citep{Tsuneta1998,Guo2012,Li2013}. The large nonthermal width of Fe~{\sc{xxi}}~could be
caused by the divergence of flows along the line of sight above the loop tops, or post-reconnection turbulence
\citep[e.g.,][]{Ciaravella2008,Guo2012,Doschek2014}. Observations by the Gamma-ray Burst Monitor (GBM) onboad the Fermi Gamma-ray Space
Telescope reveal an enhancement of hard X-rays up to 30-50 keV around 17:17:45 (Figure~\ref{fig.5}(C)), indicating electron acceleration mostly
at the beginning of the third episode of this complex flare and before the plasmoid ejection.

Some efforts have been made to search for simultaneous observations of reconnection inflows and outflows. Reconnection rates have been estimated
based on imaging observations of several flares \citep{Li2009,Cheng2010,Takasao2012,Su2013} and they are mostly in the range of 0.05-0.5. A
spectroscopic investigation by \cite{Hara2011} yielded a reconnection rate of 0.05-0.1 in a flare. Without knowing the three-dimensional
geometry of the flare, it is difficult to estimate the impact of the projection effect on the measured flow velocities. As a rough estimate,
using 20~km~s$^{-1}$ (40/2) as the inflow speed and 125~km~s$^{-1}$ as the outflow speed leads to a reconnection rate of 0.16.

\section{Flows associated with the loop legs}

\subsection{Downward-propagating blobs}

From the AIA~211\AA{}~movie we find fast downward-moving blobs along post-reconnection loops around 17:19:35. The plane of sky component of
the velocity of one blob is estimated to be $\sim$135~km~s$^{-1}$ from the space-time map of AIA~211\AA{}
(Figure~\ref{fig.3}(H)). This moving blob was captured by the IRIS slit, and we find a greatly enhanced redshifted component of the
Si~{\sc{iv}}~line profile, indicating a velocity of $\sim$60~km~s$^{-1}$ in the line of sight (location 3 in Figures~\ref{fig.3}). The full
speed can then be estimated as $\sim$150~km~s$^{-1}$, which is comparable to the observed reconnection outflow speed in the line of sight.

While outward-moving blobs have been frequently reported \citep{Ko2003,Lin2005,Savage2010,Song2012}, inward-moving blobs are rarely observed in
flares. \cite{Savage2012b} found that similar density enhancements along loop legs move even faster than the loop shrinks. These moving blobs
could be accelerated plasma from the reconnection site.

\subsection{Downward-moving loops}

Downward motion of loop legs was observed below the ejecta and the linear bright structure from 17:17:35 to 17:18:11 in the 1400\AA{}~and
211\AA{}~images. These relatively small loops (North-South orientation) may also be identified in AIA 131\AA{}~images, in which they seem to be
partly obscured by longer loops (East-West orientation) which were only present in the hotter AIA 131\AA{}~and 94\AA{}~passbands. One likely
scenario is: while one ribbon of the flare was concentrated around the coordinate of (740$^{\prime\prime}$, -207$^{\prime\prime}$), the other
ribbon was well separated, at locations of (740$^{\prime\prime}$, -185$^{\prime\prime}$) and (695$^{\prime\prime}$, -190$^{\prime\prime}$)
respectively.

The plane of sky component of the motion of these loops has been found to be $\sim$55~km~s$^{-1}$ from the AIA~211\AA{} observation
(Figure~\ref{fig.1}(I)). The downward motion of these loops may be related to the reconnection inflow, or shrinking loops resulting from a reconnection process in the early stage of this flare. Another possibility is magnetic implosion, and the contraction motion is caused by the reduction of the magnetic pressure due to energy release \citep{Hudson2000} at the beginning of the
flare.

The IRIS slit moved to the location of these loops at 17:19:03 and 17:19:35 (Figures~\ref{fig.2} and \ref{fig.3}). Interestingly, O~{\sc{iv}}, Si~{\sc{iv}}, C~{\sc{ii}} and Mg~{\sc{ii}} lines, formed well below 1 MK, all exhibited clear splitting at these locations. We clearly see a broad redshifted (by $\sim$55~km~s$^{-1}$) component besides a narrow background component. At first sight this redshifted component seems to be associated with the loop motion. However, from the high-cadence AIA 211\AA{}~movie there appeared to be no obvious loop contraction at these times. Instead, the redshifted component might be associated with falling plasma in these loops and the downward-propagating blobs are perhaps the denser part of the falling plasma.

\section{Flows at the loop footpoints}

Location 2 in Figure~\ref{fig.3} corresponds to one footpoint of the flare loops. There we find significant enhancement at the blue side of the
Fe~{\sc{xxi}}~1354.08\AA{}~rest position. Several neutral and singly ionized lines marked in Figure~\ref{fig.2}(E) contributed to this emission.
These lines, however, are narrow and cannot explain the bulk enhancement. This emission feature is likely due to the highly blueshifted (order
of $\sim$260~km~s$^{-1}$) Fe~{\sc{xxi}}~line.  At the same time, cool lines formed in the transition region and upper chromosphere (e.g.,
O~{\sc{iv}}, Si~{\sc{iv}}, C~{\sc{ii}} and Mg~{\sc{ii}}) all exhibited significant redshifts at the same location. For example, the Si~{\sc{iv}}
line is redshifted by $\sim$50~km~s$^{-1}$. This is consistent with the scenario of explosive chromospheric evaporation \citep{Fisher1985}. Using the O~{\sc{iv}}~line pair, the
electron density is found to be on the order of log ({\it N}$_{e}$/cm$^{-3}$)=11.0 at location 2. We notice that strong evaporation flows were also present at
17:18:32.

At 17:20:06 the large blueshift at location 2 can still be seen (Figure~\ref{fig.4}). A single Gaussian fit to this emission feature gives a blue shift of
$\sim$260~km~s$^{-1}$ and a nonthermal width of $\sim$44~km~s$^{-1}$. All the neutral and singly ionized lines marked in Figure~\ref{fig.2}(E)
now disappeared, perhaps because there was no chromospheric material in the higher part of the loop.

The Fe~{\sc{xxi}} line profile in the evaporation flow seems to be entirely blueshifted by $\sim$260~km~s$^{-1}$, in agreement with hydrodynamic
simulations of \cite{Liu2009}. Entirely shifted Fe~{\sc{xxi}} line profiles have also been found by \cite{Young2014} and \cite{Li2014}. This contrasts with previous observations of emission lines with formation temperatures lager than 10 MK by the
EUV Imaging Spectrometer \citep[EIS,][]{Culhane2007} onboard HINODE, where usually a stationary emission component and a blueshifted component
(or blue wing enhancement) co-exist during chromospheric evaporation \citep[e.g.,][]{Milligan2009,Li2011,Young2013}. This difference possibly
suggests that the resolution of EIS is not high enough to separate the evaporation flow from the ambient stationary hot plasma.

\section{Summary}

\begin{figure*}
\centering {\includegraphics[width=\textwidth]{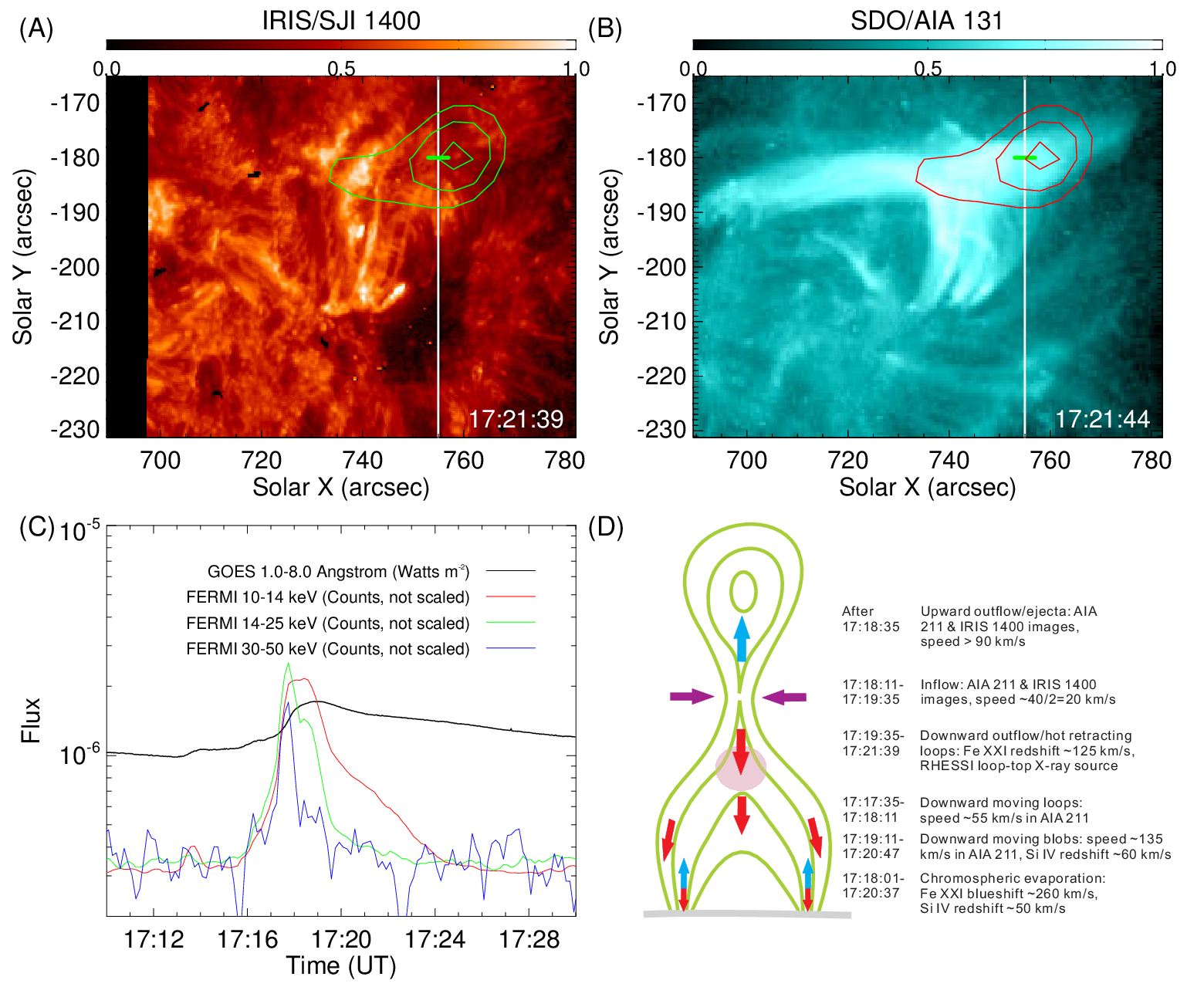}} \caption{ (A)-(B): IRIS/SJI~1400\AA{}~and 131\AA{}~images taken around 17:21:39 UT. The
white line in each panel indicates the slit location at the corresponding time. The short horizontal line marks the location where the
redshifted Fe~{\sc{xxi}}~feature was observed. Contours of RHESSI 6-12~keV X-Ray flux, which was reconstructed with the CLEAN algorithm using
detectors 3-9, are overplotted. (C): GOES and FERMI (from detector n0) fluxes in different energy bands. (D): A cartoon showing various flows
observed in this flare. } \label{fig.5}
\end{figure*}

The different types of flows observed from combined imaging and spectroscopic observations are summarized in Figures~\ref{fig.5}(D). To our
knowledge, this is one of the most complete observations of a variety of flows in one flare. The most interesting results are the following: (1)
Large red shift of Fe~{\sc{xxi}} has been observed with IRIS for the first time and interpreted as a signature of reconnection downflows; (2)
Fast blob-like structures have been observed to propagate from the reconnection site downward to the loop footpoints; (3) The Fe~{\sc{xxi}} line
profile in the evaporation flow has been observed to be entirely blueshifted by $\sim$260~km~s$^{-1}$.

\begin{acknowledgements}
IRIS is a NASA small explorer mission developed and operated by LMSAL with mission operations executed at NASA Ames Research center and major
contributions to downlink communications funded by the Norwegian Space Center (NSC, Norway) through an ESA PRODEX contract. This work is
supported by contracts 8100002705 and SP02H1701R from LMSAL to SAO, NASA grant NNX11AB61G, and NSF ATM-08477719. We thank B. Zhang for interpreting the Fermi data and L.-H. Wang for discussions.
\end{acknowledgements}

\end{document}